\documentclass[letterpaper, conference]{IEEEtran}
\newif\ifpreprint
\preprinttrue

\usepackage{cite}
\usepackage{microtype}
\usepackage{graphicx}
\usepackage{subcaption}
\usepackage{booktabs}
\usepackage{amsmath,amssymb,amsfonts}
\usepackage{mathtools}
\usepackage{algorithmic}
\usepackage{tikz}
\usepackage{standalone}
\usepackage{pgfplots}
\usepackage{textcomp}
\usepackage{xcolor}
\usepackage{dsfont}
\usepackage{bm}
\usepackage{url}
\usepackage{placeins}
\usepackage[capitalize]{cleveref}
\usepackage{mleftright}

\mleftright
\pgfplotsset{compat=newest}
\usetikzlibrary{shapes, arrows, positioning, calc}
\pgfplotstableread{data/bound_results.csv}\loadedtable
\newcommand{\<}{\negmedspace{}}

\renewcommand{\Pr}{\operatorname{Pr}}

\newcommand{\mc}{\mathcal}
\DeclareMathOperator{\E}{\mathbb{E}}
\DeclareMathOperator{\1}{\mathds{1}}
\DeclareMathOperator{\Cov}{cov}
\DeclareMathOperator{\Var}{Var}

\DeclarePairedDelimiter\ceil{\lceil}{\rceil}

\DeclareMathOperator*{\argmin}{arg\,min}

\newtheorem{theorem}{Theorem}
\newtheorem{lemma}{Lemma}

\newtheorem{corollary}{Corollary}

\begin{document}
\title{One-Shot Broadcast Joint Source-Channel Coding with Codebook Diversity}

\author{
    \IEEEauthorblockN{Joseph Rowan, Buu Phan and Ashish J. Khisti}
    \IEEEauthorblockA{Department of Electrical and Computer Engineering, University of Toronto, Toronto, Canada\\
    Email: joseph.rowan@mail.utoronto.ca, truong.phan@mail.utoronto.ca, akhisti@ece.utoronto.ca}
}

\maketitle

\begin{abstract}
We study a one-shot joint source-channel coding setting where the source is encoded once and broadcast to $\bm{K}$ decoders through independent channels.
Success is predicated on at least one decoder recovering the source within a maximum distortion constraint.
We find that in the one-shot regime, utilizing disjoint codebooks at each decoder yields a \emph{codebook diversity gain}, distinct from the \emph{channel diversity gain} that may be expected when several decoders observe independent realizations of the channel's output but share the same codebook.
Coding schemes are introduced that leverage this phenomenon, where first- and second-order achievability bounds are derived via an adaptation of the Poisson matching lemma which allows for multiple decoders using disjoint codebooks.
We further propose a hybrid coding scheme that partitions decoders into groups to optimally balance codebook and channel diversity.
Numerical results on the binary symmetric channel demonstrate that the hybrid approach outperforms strategies where the decoders' codebooks are either fully shared or disjoint.
\end{abstract}


\section{Introduction}\label{sec:intro}
In the classical joint source-channel coding (JSCC) setting \cite{csiszar1982,kostina2013}, the encoder views a source symbol $W$ and maps it to a message $X$, which is then sent over a communication channel.
The decoder's role is to observe the channel output and produce a reconstruction $\hat{Z}$, with success being contingent on recovering the source subject to a maximum distortion constraint $D$.
In this work, we are interested in a one-shot broadcast variant of JSCC where the source is encoded once and then sent simultaneously to $K$ decoders through independent channels; success requires at least one decoder to recover the source to the desired fidelity.
Moreover, each decoder may also have access to conditionally independent side information.

As one example of a potential application, consider a distributed machine learning-based sensing system in which success requires at least one sensor to register some fact about the environment, such as a nearby vehicle or trespasser, after running a downstream learning task.
Such a setting reflects standard notions of detection probability in distributed detection theory \cite{viswanathan1997,tenney1980}.
In this context, the encoder would transmit global knowledge, perhaps including a compressed representation of the entire surroundings in the form of latent variables, and the side information could consist of local details that are helpful in making the final decision.
Each decoder then runs its task, for example evaluating a trained neural network, to generate a reconstruction $\hat{Z}_k$ and uses this to decide whether to raise the alarm.
While the downstream details of the detection and machine learning system may vary, the probability of excess distortion can serve as a proxy for its reliability, with low distortion being a prerequisite for end-to-end performance.

The \emph{Poisson matching lemma} (PML) framework introduced in \cite{li2021} has recently been used to establish various one-shot achievability results, including for JSCC \cite[Th.~4]{li2021}.
Yet despite its versatility, the PML does not natively handle the specific multi-decoder setting that we tackle here.
While the \emph{generalized PML} \cite[Lem.~3]{li2021} is conceptually related in that it allows for multiple attempts to recover the source, it applies more naturally to scenarios where one decoder returns the $K$ most promising outputs based on the order statistics of a single shared Poisson process, e.g.\ as could be done in list decoding for channel coding.
In contrast, we deal with $K$ decoders operating without inter-decoder communication.

A key insight arising from the present work is that in the one-shot regime, using multiple decoders can unlock a \emph{codebook diversity gain} if we force the codebooks to be disjoint, an advantage which does not appear at long blocklengths.
Via an extension of the PML which we call the \emph{list PML}, we derive coding schemes and achievable error bounds that make use of codebook diversity gain both with and without side information.
Conversely, sharing one codebook between decoders enables a more familiar \emph{channel diversity gain} by virtue of having multiple independent looks through the channel.
We conclude by proposing a hybrid approach that balances these two opposing strategies to optimize the achievable rate, accompanied by numerical analysis of the schemes over the binary symmetric channel (BSC).
\ifpreprint\else
An extended version of the paper is provided in \cite{extendedversion}.
\fi

\section{Problem Setup}\label{sec:problem_setup}
We consider a one-shot setting with one encoder and $K$ decoders, where each decoder is connected to the encoder through a private and independent communication channel which maps an input alphabet $\mc{X}$ to an output alphabet $\mc{Y}$.
Furthermore, we restrict our attention to the case where all the channels are identically modeled by a conditional distribution $P_{Y \mid X}$.
Under this assumption, the encoder, defined by the mapping $f : \mc{W} \to \mc{X}$, observes the source symbol $W \in \mc{W}$ from $P_W$ and broadcasts a message $X \in \mc{X}$ over each channel.
Subsequently, decoder $k$ views the sample $Y_k \in \mc{Y}$ at the output of its channel and generates the reconstruction $\hat{Z}_k \in \mc{Z}$ using a mapping $g : \mc{Y} \to \mc{Z}$.
The probability of excess distortion for an individual decoder under the distortion measure $d : \mc{W} \times \mc{Z} \to \mathbb{R}_{\geq 0}$ is $P_{ek} = \Pr[d(W, \hat{Z}_k) > D]$.
As we require at least one decoder to recover the source within the specified distortion, the ensemble error probability is then $P_e = \Pr[\cap_{k=1}^{K} \{ d(W, \hat{Z}_k) > D \}]$.

\begin{figure}[tb]
    \centering
    \begin{tikzpicture}[
    node distance=1.5cm,
    block/.style={draw, rectangle, minimum height=2.5em, minimum width=3em},
    >=latex
]

\coordinate (source) {};
\node [block, right=0.8cm of source] (encoder) {Encoder $f$};
\coordinate [right=0.8cm of encoder] (branch) {}; 

% Decoder 1 Branch (Top)
\node [block, above right=0.4cm and 0.6cm of branch] (chan1) {$P_{Y \mid X}$};
\node [block, right=1.2cm of chan1, minimum width=1.4cm] (dec1) {Dec. $1$};
\node [above left=0.6cm and 0cm of dec1.west] (side1) {$T_1$};
\coordinate [right=0.8cm of dec1] (out1) {};
    
% Decoder K Branch (Bottom)
\node [block, below right=0.4cm and 0.6cm of branch] (chank) {$P_{Y \mid X}$};
\node [block, right=1.2cm of chank, minimum width=1.4cm] (deck) {Dec. $K$};
\node [above left=0.6cm and 0cm of deck.west] (sidek) {$T_K$};
\coordinate [right=0.8cm of deck] (outk) {};
    
% Ellipsis for K decoders
\path (chan1) -- (chank) node [midway, yshift=3pt] {$\vdots$};
\path (dec1) -- (deck) node [midway, yshift=3pt] {$\vdots$};

% Connections
\draw [->] (source) node[left] {$W$} -- (encoder);
\draw (encoder) -- node[above] {$X$} (branch);
    
% Connect to channels
\draw [->] (branch.center) |- (chan1);
\draw [->] (branch.center) |- (chank);
    
% Connect channels to decoders
\draw [->] ([yshift=-5pt]chan1.east) node[above, anchor=south west] {$Y_1$} -- ([yshift=-5pt]dec1.west);
\draw [->] ([yshift=-5pt]chank.east) node[above, anchor=south west] {$Y_K$} -- ([yshift=-5pt]deck.west);
    
% Side Information
\draw [->] (side1) |- ([yshift=5pt]dec1.west);
\draw [->] (sidek) |- ([yshift=5pt]deck.west);

% Outputs
\draw [->] (dec1) -- (out1) node[right] {$\hat{Z}_1$};
\draw [->] (deck) -- (outk) node[right] {$\hat{Z}_K$};

\end{tikzpicture}
    \captionsetup{belowskip=-1.3em}
    \caption{System model for broadcast JSCC with side information. The source $W$ is encoded to $X$. Decoder $k$ uses the channel output $Y_k$ and side information $T_k$ to produce its reconstruction $\hat{Z}_k$.}
    \label{fig:system_diagram}
\end{figure}
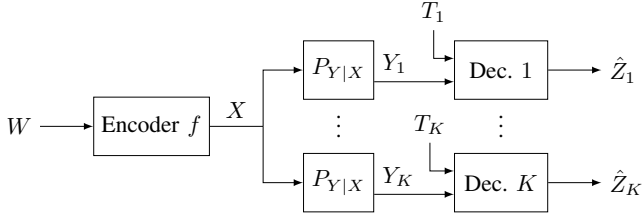

We further consider a setting where each decoder has access to a conditionally iid side information realization $T_k \in \mc{T}$ from $P_{T \mid W}$, as motivated in \cref{sec:intro}.
As opposed to the basic setting, decoder $k$ now generates $\hat{Z}_k$ using a mapping $g : \mc{Y} \times \mc{T} \to \mc{Z}$ that can exploit the side information as well as the channel observation.
This constitutes a variant of the Wyner-Ziv problem \cite{wyner1976} and consequently requires a somewhat different analysis.
A block diagram of the full system model with side information is illustrated in \cref{fig:system_diagram}.

In what follows, we assume that all alphabets, e.g.\ $\mc{W}$, $\mc{X}$ and $\mc{Y}$, are standard Borel spaces endowed with their respective Borel $\sigma$-algebras and assume a common $\sigma$-finite reference measure $\theta$ on each space.
For a pair of measures $\mu$, $\nu$ on $\mc{U}$ such that $\nu$ is absolutely continuous with respect to $\mu$, denoted as $\nu \ll \mu$, the Radon-Nikodym derivative is $d\nu / d\mu : \mathcal{U} \to \mathbb{R}_{\geq 0}$.
To simplify the exposition, we assume that all probability measures to be discussed are absolutely continuous with respect to the reference measure $\theta$ on the appropriate space and work in terms of the density function $p(u) = (dP / d\theta)(u)$ for any measure $P$.
Furthermore, we write $\lambda$ for the Lebesgue measure on $\mathbb{R}$ and $\lambda_S$ for its restriction to some $S \subseteq \mathbb{R}$.
Logarithms are base-$2$ throughout; the information density for a pair of random variables $(X,Y) \sim P_{X,Y}$ is defined as
\begin{IEEEeqnarray}{c}
    \iota_{X;Y}(x;y) = \log \frac{p_{X,Y}(x,y)}{p_X(x) p_Y(y)} = \log \frac{p_{Y \mid X}(y \mid x)}{p_Y(y)}.
\end{IEEEeqnarray}

\section{Main Results}\label{sec:main_results}
\subsection{One-Shot Achievability with Disjoint Codebooks}
In this section, we present one-shot achievability results for the broadcast JSCC problem with codebook diversity introduced in \cref{sec:problem_setup}, first considering the simpler setting without side information.
While the channel statistics $P_{Y \mid X}$ are assumed to be the same for each of the $K$ receivers, the realization $Y_k$ arriving at a given receiver is taken to be conditionally independent of the others' observations given the input $X$.

\begin{theorem}\label{thm:oneshot_jscc}
Fix $P_X$ and $P_Z$.
Then, with $P_W$ being the source distribution and $P_{Y \mid X}$ the channel, there exists a code for $K$ decoders with error probability satisfying
\begin{IEEEeqnarray}{c}
    P_e \leq \E\bigl[ \bigl( 1 + K P_Z(\mathcal{B}_D(W)) 2^{\iota_{X;Y}(X;Y)} \bigr)^{-1} \bigr] \label{eqn:oneshot_jscc}
\end{IEEEeqnarray}
where $\mathcal{B}_D(w) = \{ z \in \mc{Z} : d(w, z) \leq D \}$ is the permissible distortion ball and $(W, X, Y) \sim P_W \times P_X P_{Y \mid X}$.
\end{theorem}

To allow for the use of side information at each decoder, we now state a variant of \cref{thm:oneshot_jscc} that integrates the Wyner-Ziv coding framework in a similar fashion to \cite[Th.~3]{li2021}.

\begin{theorem}\label{thm:oneshot_jscc_wz}
Fix $P_X$, a test channel $P_{U|W}$ and a reconstruction function $\phi : \mc{U} \times \mc{T} \to \mc{Z}$.
Then, with $P_W$ being the source distribution and $P_{Y \mid X}$ the channel, there exists a code for $K$ decoders with error probability satisfying
\begin{IEEEeqnarray}{c}
    P_e \leq \E\bigl[ \bigl( 1 + K \1\{ d(W, \phi(U, T)) \! \leq \! D \} \, 2^{S(W, U, T, X, Y)} \bigr)^{-1} \bigr] \IEEEeqnarraynumspace \label{eqn:oneshot_jscc_wz}
\end{IEEEeqnarray}
where $(W, U, T, X, Y) \sim P_W P_{U \mid W} P_{T \mid W} \times P_X P_{Y \mid X}$ and $S(W, U, T, X, Y) = \iota_{X;Y}(X;Y) + \iota_{U;T}(U;T) - \iota_{U;W}(U;W)$.
\end{theorem}

Compared to the similar single-user results in \cite[Th.~4]{li2021} and \cite[Th.~3]{li2021} for one-shot JSCC and Wyner-Ziv coding respectively, the codebook diversity gain in \cref{thm:oneshot_jscc,thm:oneshot_jscc_wz} appears as a factor of $K$ in the denominators of \eqref{eqn:oneshot_jscc} and \eqref{eqn:oneshot_jscc_wz}.
Since the gain arises solely due to the use of $K$ disjoint sub-codebooks, it does not depend on the channel or source statistics.
We will return to specify the coding strategy, which we call the \emph{disjoint scheme}, and achievability proofs in \cref{sec:jscc_proof,sec:jscc_wz_proof}, as this requires an extension of the PML allowing the error probability to be bounded in the desired multi-decoder setting.
Such an extension, the list PML, will be presented in \cref{sec:lemmas}.

\subsection{Second-Order Analysis}\label{sec:second_order_analysis}
We state a second-order result on the achievable rate for the disjoint-codebook setup without side information in \cref{thm:oneshot_jscc}.
Following \cite[Prop.~5]{li2021}, assume an iid source sequence $W^m$, distortion measure $d(w^m, \hat{z}^m) = \sum_{i=1}^{m} d(w_i, \hat{z}_i) / m$ and $n$ uses of the channel $P_{Y \mid X}$.
The rate-distortion function is
\begin{IEEEeqnarray}{c}
    R(D) = \inf_{P_{Z \mid W} : \E[d(W, Z)] \leq D} I(W; Z)
\end{IEEEeqnarray}
and we assume $P_{Z \mid W}$ attains the optimal rate for distortion $D$.
Let $P_Z$ be the marginal of $P_W P_{Z \mid W}$.
The $D$-tilted information \cite{kostina2012} is then defined as $\jmath_W(w, D) = -{\log}\E_{Z \sim P_Z}[2^{-(D - d(w, Z)) R'(D)}]$.
The following theorem relies on \cite[Lem.~2]{kostina2012} concerning the second-order rates of  lossy source coding, which in turn imposes additional technical conditions on the quantities just introduced.
These conditions are detailed in \ifpreprint \cref{sec:dispersion_proof}; \else \cite[App.~D]{extendedversion}; \fi we assume in what follows that all are satisfied.

\begin{theorem}\label{thm:oneshot_jscc_dispersion}
For $0 < \varepsilon < 1$ and all sufficiently large $m$, $n$, the error probability satisfies $P_e \leq \varepsilon$ if
\begin{IEEEeqnarray}{l}
    nC - mR(D) \geq \alpha \log m + \frac{1}{2} \log n + \beta - \log K \nonumber \\
    \IEEEeqnarraymulticol{1}{r}{
        \qquad \< + \sqrt{nV + m \mathcal{V}(D)} Q^{-1}\biggl( \varepsilon - \frac{\eta}{\sqrt{\min\{ n, m \}}} \biggr)
    } \label{eqn:jscc_dispersion_condition}
\end{IEEEeqnarray}
where $C = I(X;Y)$ is the channel capacity, $V = \Var(\iota_{X;Y}(X;Y))$ is the channel dispersion, $\mc{V}(D) = \Var(\jmath_W(W,D))$ is the source dispersion, $\alpha,\beta > 0$ are constants given by \cite[Lem.~2]{kostina2012} and $\eta > 0$ is a constant given by the Berry-Esseen theorem \cite{feller1971}.
\end{theorem}

The proof is presented in \ifpreprint \cref{sec:dispersion_proof} \else \cite[App.~D]{extendedversion} \fi and involves applying the Berry-Esseen theorem to the bound in \eqref{eqn:oneshot_jscc}, similar to the approach used to prove \cite[Prop.~5]{li2021}.
Compared to the result obtained there for point-to-point JSCC, the availability of multiple decoders using separate codebooks in the disjoint scheme provides a gain proportional to $\log K / n$ in the achievable rate.
While this advantage disappears as $n$ increases, it can be significant for short blocklengths or large $K$.

\subsection{Specialization to Near-Lossless JSCC}
\Cref{thm:oneshot_jscc,thm:oneshot_jscc_wz} also admit simpler and more intuitive forms in the case of near-lossless JSCC with a discrete source \cite{campo2011,kostina2013}.
We deal with this special case in \cref{cor:near_lossless_jscc} below.

\begin{corollary}\label{cor:near_lossless_jscc}
Let $W$ be drawn from a discrete alphabet $\mathcal{W}$ and the distortion measure be the Hamming distortion $d(w, z) = \1\{ w \neq z \}$ with $D=0$.
Then:
\begin{enumerate}
    \item By choosing $P_Z = P_W$ in \cref{thm:oneshot_jscc}, the error probability satisfies
    \begin{IEEEeqnarray}{c}
        P_e \leq \E\bigl[ \bigl( 1 + K P_W(W) 2^{\iota_{X;Y}(X;Y)} \bigr)^{-1} \bigr]. \label{eqn:corr1}
    \end{IEEEeqnarray}
    If $W$ is uniform on a set of cardinality $M$, this reduces to $P_e \leq \E[ (1 + K M^{-1} 2^{\iota_{X;Y}(X;Y)})^{-1} ]$, recovering a standard channel coding bound analogous to \cite[Prop.~1]{li2021}.

    \item By fixing $U = W$ and choosing $\phi(U, T) = U$ in \cref{thm:oneshot_jscc_wz}, the bound simplifies to
    \begin{IEEEeqnarray}{c}
        P_e \leq \E\bigl[ \bigl( 1 + K P_{W \mid T}(W \mid T) 2^{\iota_{X;Y}(X;Y)} \bigr)^{-1} \bigr]
    \end{IEEEeqnarray}
    representing a one-shot Slepian-Wolf-type result \cite{slepian1973} where the source is characterized by the posterior $P_{W \mid T}$.
\end{enumerate}
\end{corollary}

\subsection{Baseline Coding Scheme and Analysis}
In this section, we introduce a baseline approach that leverages channel diversity by applying a single-user PML coding strategy to the broadcast setting with $K$ decoders.
We focus on the problem as considered in \cref{thm:oneshot_jscc} where side information is not available.
In the baseline scheme, the single-user JSCC method from \cite[Th.~4]{li2021} is repeated $K$ times using the same codebook; the codebook generation, encoding and decoding procedures to be outlined in \cref{sec:jscc_proof} are followed except that decoder $k$ instead chooses the index
\begin{IEEEeqnarray}{c}
    K_{qk} = \argmin_{i \geq 1} \frac{\tau_i p_X(X_i) p_Z(Z_i)}{p_{X \mid Y}(X_i \mid Y_k) p_Z(Z_i)}.
\end{IEEEeqnarray}
Mirroring the approach that will be used to prove \cref{thm:oneshot_jscc}, we show in \ifpreprint \cref{sec:hybrid_proof} \else \cite[App.~C]{extendedversion} \fi that the baseline achieves
\begin{IEEEeqnarray}{c}
    P_e \leq \E\Bigl[ \Bigl( 1 + P_Z(\mathcal{B}_D(W)) \max_{1 \leq k \leq K} 2^{\iota_{X;Y}(X;Y_k)} \Bigr)^{-1} \Bigr]. \IEEEeqnarraynumspace \label{eqn:baseline_bound}
\end{IEEEeqnarray}
This scheme maximally exploits diversity in the channel outputs, which is reflected in the fact that the information density term in \eqref{eqn:baseline_bound} is computed with respect to the single best observation, i.e.\ that for which $\iota_{X;Y}(X;Y_k)$ is maximized, rather than the average channel statistics.
In contrast, the codebook diversity gain afforded by our disjoint scheme in \cref{thm:oneshot_jscc} introduces a factor of $K$ in \eqref{eqn:oneshot_jscc} by partitioning the codebook.
This latter strategy is favorable when either $K$ is large, the blocklength is short or the channel SNR is high; neither approach is universally better than the other.
As all decoders search the same codebook in the baseline approach, it may also be used in situations where $1 < K' \leq K$ decoders must succeed, which is not possible with the disjoint scheme.
Since success at each decoder is not independent, obtaining a tight bound in this setting requires a different analysis, which we leave to future work.

\subsection{Hybrid Strategy}
We present a hybrid approach that combines the advantages of both schemes, again restricting our attention to the case without side information.
Split the $K$ decoders into $J$ groups of size $L = K / J$.
The codebook is partitioned between the groups, while the decoders in a group share the same sub-codebook.
We re-use the procedure in \cref{sec:jscc_proof} except that each $C_i$ is now uniform on $\{ 1, \ldots, J \}$ and decoder $k$ chooses
\begin{IEEEeqnarray}{c}
    K_{qk} = \argmin_{i \geq 1, C_i = \ceil{k / L}} \frac{\tau_i p_X(X_i) p_Z(Z_i)}{p_{X \mid Y}(X_i \mid Y_k) p_Z(Z_i)}. \label{eqn:hybrid_decoder_selection}
\end{IEEEeqnarray}
As shown in \ifpreprint \cref{sec:hybrid_proof}, \else \cite[App.~C]{extendedversion}, \fi the error probability satisfies
\begin{IEEEeqnarray}{c}
    P_e \leq \E\Bigl[ \Bigl( 1 + J P_Z(\mc{B}_D(W)) \max_{1 \leq k \leq L} 2^{\iota_{X;Y}(X; Y_k)} \Bigr)^{-1} \Bigr]. \IEEEeqnarraynumspace \label{eqn:hybrid_bound}
\end{IEEEeqnarray}
This strategy recovers the disjoint scheme when $J = K$ and the baseline when $J = 1$.
In between these extremes, which maximize the codebook diversity and channel diversity gains respectively, $J$ can be chosen to balance the two gains via a 1D search.
Furthermore, we have the following second-order result without side information, assuming the conditions required by \cref{thm:oneshot_jscc_dispersion} are satisfied.

\begin{theorem}\label{thm:hybrid_second_order}
For $0 < \varepsilon < 1$ and all sufficiently large $m$, $n$, the hybrid scheme with $J$ groups of size $L$ achieves $P_e \leq \varepsilon$ if
\begin{IEEEeqnarray}{l}
    nC - mR(D) \geq \alpha \log m + \frac{1}{2} \log n + \beta - \log J \nonumber \\
    \IEEEeqnarraymulticol{1}{r}{
        \qquad \< - \sqrt{n + m} F^{-1}\biggl( \varepsilon - \frac{\eta}{( \min\{ n, m \} )^{ \min\{ c, 1/2 \}}}; \Sigma \biggr) \IEEEeqnarraynumspace
    } \label{eqn:hybrid_dispersion_condition}
\end{IEEEeqnarray}
where $F^{-1}(t; \Sigma)$ is the quantile function of the maximum coordinate of an $L$-dimensional Gaussian random vector with mean zero and covariance matrix $\Sigma = (\sigma_1^2 1_{L \times L} + \sigma_2^2 I) / (n + m)$, $1_{L \times L}$ is an $L \times L$ matrix of ones, $\sigma_1^2 = n \tilde{V} + m \mathcal{V}(D)$, $\sigma_2^2 = n (V - \tilde{V})$, $\tilde{V} = \Var(D_{\mathrm{KL}}(P_{Y \mid X}(\cdot \mid X) \Vert P_Y))$, $\eta, c > 0$ are constants given by the Chernozhukov-Chetverikov-Kato bound \cite{chernozhukov2013} and all other quantities are defined as in \cref{thm:oneshot_jscc_dispersion}.
\end{theorem}

The proof is deferred to \ifpreprint \cref{sec:hybrid_dispersion_proof}. \else \cite[App.~E]{extendedversion}. \fi
As opposed to \cref{thm:oneshot_jscc_dispersion}, which leverages the one-dimensional Berry-Esseen theorem \cite{feller1971}, \cref{thm:hybrid_second_order} relies on a Gaussian approximation of the maximum coordinate of a sum of random vectors.

\section{List Poisson Matching Lemma}\label{sec:lemmas}
We now present an adaptation of the standard PML that will be used to prove the results in \cref{sec:main_results}.
The motivating idea behind our extension is to move from a point-wise matching framework with one global codebook to a setting where the codebook is partitioned between $K$ decoders, allowing multiple chances to obtain the correct index.
Again, we stress that this is different from the generalized PML \cite{li2021}, in which a single decoder with access to the entire codebook outputs several different guesses based on order statistics.
\begin{lemma}[List PML]\label{lemma:clml}
Let $\Pi = \{ U_i, \tau_i \}_{i \in \mathbb{N}}$ be a marked Poisson point process on $\mathcal{U} \times \mathbb{R}_{\geq 0}$ with intensity measure $\mu \times \lambda_{\mathbb{R}_{\geq 0}}$.
Further associate with each point of the process an independent discrete random variable $Z_i$ uniformly distributed on $\{ 1, \ldots, K \}$ and let $P$, $Q \ll \mu$ be two probability measures on $\mc{U}$.
Select the indices
\begin{IEEEeqnarray}{c}
    K_p = \argmin_{i \geq 1} \frac{\tau_i \mu(U_i)}{p(U_i)}, \ K_{qk} = \argmin_{i \geq 1, Z_i = k} \frac{\tau_i \mu(U_i)}{q(U_i)}
\end{IEEEeqnarray}
and define $U_p = U_{K_p}$, $Z_p = Z_{K_p}$ and $U_{qk} = U_{K_{qk}}$.
We have
\begin{IEEEeqnarray}{c}
    \Pr[U_p \notin \{ U_{qk} \}_{k=1}^{K} \mid U_p] \leq 1 - \biggl( 1 + \frac{p(U_p)}{K q(U_p)} \biggr)^{-1}. \label{eqn:clml}
\end{IEEEeqnarray}
\end{lemma}

We provide a proof in \ifpreprint \cref{sec:lml_proof}. \else \cite[App.~A]{extendedversion}. \fi
By assigning the random label $Z_i \in \{ 1, \ldots, K \}$ to each point in the Poisson point process $\Pi$, we are able to invoke the thinning property \cite{last2017} to split the process into $K$ independent sub-processes, each with intensity measure $(\mu \times \lambda_{\mathbb{R}_{\geq 0}}) / K$.
In the coding schemes to be introduced in \cref{sec:jscc_proof,sec:jscc_wz_proof}, this will allow each decoder to search for a matching codeword in a disjoint codebook that is not visible to the other decoders, thereby unlocking the codebook diversity gain that appears in \eqref{eqn:oneshot_jscc} and \eqref{eqn:oneshot_jscc_wz}.

In order to apply \cref{lemma:clml} to the broadcast JSCC problem, a conditional version is also required.

\begin{lemma}[Conditional list PML]\label{lemma:cclml}
Let the Poisson process $\Pi$ be as in \cref{lemma:clml}, and again associate an independent discrete uniform random variable $Z_i \in \{ 1, \ldots, K \}$ with each point of the process.
Fix a joint distribution $P_{X,U,Y}$ together with a probability kernel $Q_{U \mid Y}$ satisfying $P_{U \mid X}(\cdot \mid X)$, $Q_{U \mid Y}(\cdot \mid Y) \ll \mu$ almost surely, and take $X \sim P_X$ independent of $\Pi$ and each $Z_i$.
Select the index
\begin{IEEEeqnarray}{c}
    K_p = \argmin_{i \geq 1} \frac{\tau_i \mu(U_i)}{p_{U \mid X}(U_i \mid X)}
\end{IEEEeqnarray}
and set $U_p = U_{K_p}$ and $Z_p = Z_{K_p}$.
Given $U_p$, let $Y_{1:K} \sim P_{Y \mid X, U}(\cdot \mid X, U_p)$ iid.
We have $(X, U_p, Y_k) \sim P_{X,U,Y}$ for each $k$ and note that the Markov chain condition $Y_{1:K} \leftrightarrow (X, U_p) \leftrightarrow \{ U_i, \tau_i \}_{i \in \mathbb{N}}$ holds by construction.
Next, select
\begin{IEEEeqnarray}{c}
    K_{qk} = \argmin_{i \geq 1, Z_i = k} \frac{\tau_i \mu(U_i)}{q_{U \mid Y}(U_i \mid Y_k)}.
\end{IEEEeqnarray}
With $U_{qk} = U_{K_{qk}}$, we have
\begin{IEEEeqnarray}{rCl}
    \IEEEeqnarraymulticol{3}{l}{
        \Pr[U_p \notin \{ U_{qk} \}_{k=1}^{K} \mid X, U_p, \: Y_{1:K}]
    }\nonumber\\* \quad
    & \leq & 1 - \sum_{k=1}^{K}\biggl( K + \frac{p_{U \mid X}(U_p \mid X)}{q_{U \mid Y}(U_p \mid Y_k)} \biggr)^{-1}.
\end{IEEEeqnarray}
\end{lemma}

The proof is given in \ifpreprint \cref{sec:conditional_lml_proof} \else \cite[App.~B]{extendedversion} \fi and is similar to that of \cref{lemma:clml}, the key difference being that $p_{U \mid X}(\cdot \mid X)$ and $q_{U \mid Y}(\cdot \mid Y_k)$ are used in place of $p$ and $q$ respectively after partitioning over the possible outcomes of $Z_p$.

\section{Proof of \Cref{thm:oneshot_jscc}}\label{sec:jscc_proof}
Armed with the conditional list PML from \cref{lemma:cclml}, we can now detail the disjoint coding scheme that achieves the bound in \cref{thm:oneshot_jscc}.
The main idea is to generate the encoder's codebook as a Poisson point process and have each decoder operate on an independent thinning of this process.

First, we generate the full codebook independently from the source $W$ as a marked Poisson point process $\Pi = \{ (X_i, Z_i), \tau_i \}_{i \in \mathbb{N}}$ on $\mathcal{X} \times \mathcal{Z} \times \mathbb{R}_{\geq 0}$ with intensity measure $P_X \times P_Z \times \lambda_{\mathbb{R}_{\geq 0}}$, and associate with each point in the codebook a uniform label $C_i \in \{ 1, \ldots K \}$ independent of $\Pi$.
Following the construction in \cite[Th.~4]{li2021}, we define a modified conditional distribution $P_{\tilde{Z} \mid W}$ to take into account the intersection with the distortion ball $\mc{B}_D(w)$.
The conditional density function is
\begin{IEEEeqnarray}{c}
    p_{\tilde{Z} \mid W}(z \mid w) = \begin{cases}
        \1\{ z \in \mc{B}_D(w) \} p_Z(z) / \rho(w), & \hspace{-1.5ex} \rho(w) > 0 \\
        p_Z(z), & \hspace{-1.5ex} \rho(w) = 0
    \end{cases} \IEEEeqnarraynumspace \label{eqn:modified_encoder_dist}
\end{IEEEeqnarray}
where $\rho(w) = P_Z(\mc{B}_D(w))$ is the total probability mass inside the ball.
Using the source symbol $W$, the encoder chooses
\begin{equation}\label{eqn:thm1_encoder_selection}
    K_p = \argmin_{i \geq 1} \frac{\tau_i p_X(X_i) p_Z(Z_i)}{p_X(X_i) p_{\tilde{Z} \mid W}(Z_i \mid W)}.
\end{equation}
The communicated message is thus $X = X_{K_p}$, and we set $\tilde{Z} = Z_{K_p}$.
Decoder $k$ observes $Y_k \sim P_{Y \mid X}(\cdot \mid X)$ and selects
\begin{equation}\label{eqn:thm1_decoder_selection}
    K_{qk} = \argmin_{i \geq 1, \, C_i = k} \frac{\tau_i p_X(X_i) p_Z(Z_i)}{p_{X \mid Y}(X_i \mid Y_k) p_Z(Z_i)}.
\end{equation}
The reconstruction is $\hat{Z}_k = Z_{K_{qk}}$.
We also define $\hat{X}_k = X_{K_{qk}}$.
For each $k$, we have $(W, \tilde{Z}, X, Y_k) \sim P_W P_{\tilde{Z} \mid W} \times P_X P_{Y \mid X}$ by \cref{lemma:cclml}.
The probability of excess distortion satisfies
\begin{IEEEeqnarray}{rCl}
    \IEEEeqnarraymulticol{3}{l}{
        \Pr[\cap_{k=1}^{K} \{ d(W, \hat{Z}_k) > D \}]
    }\nonumber\\* \quad
    & \leq & \Pr[\rho(W) = 0] + \Pr[\rho(W) > 0, \tilde{Z} \notin \{ \hat{Z}_k \}_{k=1}^{K}] \\
    & \leq & \Pr[\rho(W) = 0] + \E[\1\{ \rho(W) > 0 \} \nonumber \\
    &&\< \times \Pr[(X, \tilde{Z}) \notin \{ (\hat{X}_k, \hat{Z}_k) \}_{k=1}^{K} \mid W, X, \tilde{Z}, Y_{1:K}]] \IEEEeqnarraynumspace \\
    & \stackrel{\text{(a)}}{\leq} & \Pr[\rho(W) = 0] + \E\biggl[\1\{ \rho(W) > 0 \} \nonumber \\
    &&\< \times \biggl( 1 - \sum_{k=1}^{K} \biggl( K + \frac{p_X(X) p_{\tilde{Z} \mid W}(\tilde{Z} \mid W)}{p_{X \mid Y}(X \mid Y_k) p_Z(\tilde{Z})} \biggr)^{-1} \biggr) \biggr] \\
    & = & \Pr[\rho(W) = 0] + \E\biggl[ \1\{ \rho(W) > 0 \} \nonumber \\
    &&\< \times \biggl(1 - \sum_{k=1}^{K} \bigl( K + \rho(W)^{-1} 2^{-\iota_{X;Y}(X;Y_k)} \bigr)^{-1} \biggr) \biggr] \\
    & \stackrel{\text{(b)}}{=} & 1 - K\E\bigl[ \bigl( K + \rho(W)^{-1} 2^{-\iota_{X; Y}(X;Y)} \bigr)^{-1} \bigr] \\
    & = & \E\bigl[ \bigl( 1 + K P_Z(\mathcal{B}_D(W)) 2^{\iota_{X;Y}(X;Y)} \bigr)^{-1} \bigr]
\end{IEEEeqnarray}
where $(W, X, Y) \sim P_W \times P_X P_{Y \mid X}$, (a) is by \cref{lemma:cclml} and (b) uses the fact that the $Y_k$'s are identically distributed given $X$ and the identity $\E[\1\{ \rho(W) > 0 \}] = \Pr[\rho(W) > 0]$. \hfill \IEEEQED{}

\section{Proof of \Cref{thm:oneshot_jscc_wz}}\label{sec:jscc_wz_proof}
We now establish the bound with side information in \cref{thm:oneshot_jscc_wz}.
Since the side information is not visible to the encoder, it cannot restrict the codeword to the permissible distortion ball as in the proof of \cref{thm:oneshot_jscc}.
Let the codebook be a marked Poisson point process $\Pi = \{ (X_i, U_i), \tau_i \}_{i \in \mathbb{N}}$ on $\mathcal{X} \times \mathcal{U} \times \mathbb{R}_{\geq 0}$ with intensity measure $P_X \times P_U \times \lambda_{\mathbb{R}_{\geq 0}}$ independent of $W$, where $P_U$ is the marginal of $P_W P_{U \mid W}$.
We associate an independent uniform label $C_i \in \{ 1, \ldots K \}$ with each point.
The encoder observes $W$ and selects
\begin{IEEEeqnarray}{c}
    K_p = \argmin_{i \geq 1} \frac{\tau_i p_X(X_i) p_U(U_i)}{p_X(X_i) p_{U \mid W}(U_i \mid W)}.
\end{IEEEeqnarray}
Let $X = X_{K_p}$ and $U = U_{K_p}$.
Upon observing $Y_k \sim P_{Y \mid X}(\cdot \mid X)$ and $T_k \sim P_{T \mid W}(\cdot \mid W)$, decoder $k$ chooses
\begin{IEEEeqnarray}{c}
    K_{qk} = \argmin_{i \geq 1, \, C_i = k} \frac{\tau_i p_X(X_i) p_U(U_i)}{p_{X \mid Y}(X_i \mid Y_k) p_{U \mid T}(U_i \mid T_k)}.
\end{IEEEeqnarray}
Let $\hat{U}_k = U_{K_{qk}}$ and $\hat{X}_k = X_{K_{qk}}$.
The reconstruction is $\hat{Z}_k = \phi(\hat{U}_k, T_k)$.
For each $k$, we have $(W, U, T_k, X, Y_k) \sim P_W P_{U \mid W} P_{T \mid W} \times P_X P_{Y \mid X}$ by \cref{lemma:cclml}.
Also, define $\mc{K}(W, U, T_{1:K}) = \{ k \in \{1, \ldots, K \} : d(W, \phi(U, T_k)) \leq D \}$ to be the set of decoders that would achieve the required distortion if $\hat{U}_k = U$.
We will refer to this set as $\mc{K}$ when the values of $W$, $U$ and $T_{1:K}$ are clear from the context.
Then,
\begin{IEEEeqnarray}{rCl}
    \IEEEeqnarraymulticol{3}{l}{
        \Pr[\cap_{k=1}^{K} \{ d(W, \hat{Z}_k) > D \}]
    }\nonumber\\* \:
    & \leq & \Pr[U \notin \{ \hat{U}_k \}_{k \in \mc{K}(W, U, T_{1:K})}] \\
    & \leq & \E[\Pr[(X, U) \notin \{ (\hat{X}_k, \hat{U}_k) \}_{k \in \mc{K}(W, U, T_{1:K})} \nonumber \\
    &&\< \mid W, X, U, Y_{1:K}, T_{1:K}]] \\
    & \stackrel{\text{(a)}}{\leq} & \E\biggl[ 1 - \sum_{k \in \mc{K}} \biggl( K + \frac{p_X(X) p_{U \mid W}(U \mid W)}{p_{X \mid Y}(X \mid Y_k) p_{U \mid T}(U \mid T_k)} \biggr)^{-1} \biggr] \\
    & = & \E\biggl[ 1 - \sum_{k = 1}^{K} \1\{ k \in \mc{K} \} \bigl( K + 2^{-S(W, U, T_k, X, Y_k)} \bigr)^{-1} \biggr] \\
    & \stackrel{\text{(b)}}{=} & \E\bigl[ \bigl( 1 + K \1\{ d(W, \phi(U, T)) \! \leq \! D \} 2^{S(W, U, T, X, Y)} \bigr)^{-1} \bigr] \IEEEeqnarraynumspace
\end{IEEEeqnarray}
where $(W, U, T, X, Y) \sim P_W P_{U \mid W} P_{T \mid W} \times P_X P_{Y \mid X}$, (a) is by \cref{lemma:cclml} and (b) follows as the $T_k$'s and $Y_k$'s are identically distributed given $W$ and $X$. \hfill \IEEEQED{}

\section{Numerical Example}\label{sec:comparison}
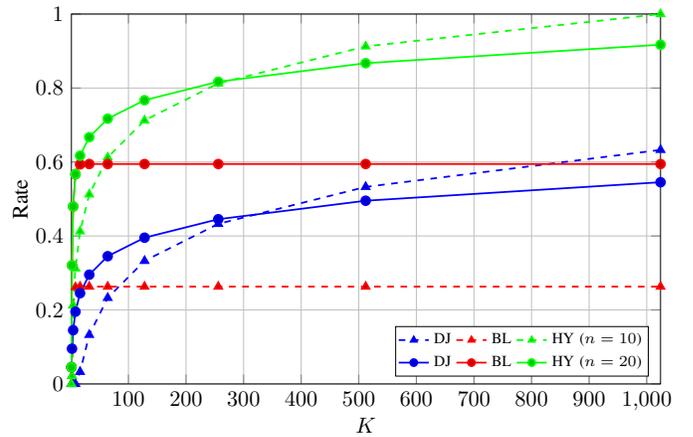
\begin{figure}[t]
    \centering
    \begin{tikzpicture}
    \begin{axis}[
        ylabel={Rate (bits)},
        xlabel={$K$},
        width=11cm, height=7.5cm,
        xmin=1, xmax=1024,
        legend style={at={(0.98,0.02)}, anchor=south east, cells={anchor=west}, inner xsep=1pt, inner ysep=1pt, font=\scriptsize},
        legend image post style={xscale=0.8, yscale=0.8},
        legend columns=3,
        xlabel near ticks,
        ylabel near ticks,
        scaled y ticks=false,
        label style={inner sep=0pt},
        yticklabel style={/pgf/number format/fixed},
        grid=major,
        ymin=0, ymax=1,
    ]
        % n = 25
        \addplot[blue,every mark/.append style={solid,fill=blue!80!black}, dashed, thick, mark=triangle*] table [x=NumDecoders, y=Rate10_Ours] {data/bound_results.csv};
        \addlegendentry{DJ}

        % n = 25
        \addplot[red,every mark/.append style={solid,fill=red!80!black}, dashed, thick, mark=triangle*] table [x=NumDecoders, y=Rate10_BL] {data/bound_results.csv};
        \addlegendentry{BL}
        
        % n = 25
        \addplot[green,every mark/.append style={solid,fill=green!80!black}, dashed, thick, mark=triangle*] table [x=NumDecoders, y=Rate10_Hybrid] {data/bound_results.csv};
        \addlegendentry{HY ($n = 10$)}
        
        % n = 50
        \addplot[blue,every mark/.append style={fill=blue!80!black}, thick, mark=*] table [x=NumDecoders, y=Rate20_Ours] {data/bound_results.csv};
        \addlegendentry{DJ}
        
        % n = 50
        \addplot[red,every mark/.append style={fill=red!80!black}, thick, mark=*] table [x=NumDecoders, y=Rate20_BL] {data/bound_results.csv};
        \addlegendentry{BL}
        
        % n = 50
        \addplot[green,every mark/.append style={fill=green!80!black}, thick, mark=*] table [x=NumDecoders, y=Rate20_Hybrid] {data/bound_results.csv};
        \addlegendentry{HY ($n = 20$)}
        
    \end{axis}
\end{tikzpicture}
    \captionsetup{belowskip=-1em}
    \caption{Comparison of achievable rates with $K$ decoders for a BSC with crossover probability $\delta = 0.05$, error tolerance $\varepsilon = 10^{-2}$ and $n \in \{10, 20\}$.
    The disjoint (DJ), baseline (BL) and hybrid (HY) schemes are shown.}
    \label{fig:bound_comparison}
\end{figure}

We now present a concrete numerical example by specializing the comparison of the coding schemes in \cref{sec:main_results} to the case of near-lossless JSCC over a BSC with crossover probability $\delta$ and blocklength $n$.
Let $W$ be uniform on $\{ 1, \ldots, M \}$, where $M = 2^{nR}$ when the rate is $R$, and set $P_X = P_W$.
Define $\bar{\delta} = 1 - \delta$.
\Cref{cor:near_lossless_jscc}, \eqref{eqn:baseline_bound} and \eqref{eqn:hybrid_bound} give
\begin{IEEEeqnarray}{rCl}
    P_{e,\mathsf{DJ}} & \leq & \sum_{t=0}^{n} \! \binom{n}{t} \delta^t \bar{\delta}^{n-t} \bigl( 1 + K M^{-1} ( \delta / \bar{\delta} )^t (2\bar{\delta})^n \bigr)^{-1} \label{eqn:coding_bound_lpml} \IEEEeqnarraynumspace \\
    P_{e,\mathsf{BL}} & \leq & \sum_{t=0}^{n} \psi_K(t; n, \delta) \bigl( 1 + M^{-1} (\delta / \bar{\delta})^t (2 \bar{\delta})^n \bigr)^{-1} \label{eqn:coding_bound_bl} \\
    P_{e,\mathsf{HY}} & \leq & \sum_{t=0}^{n} \psi_L(t; n, \delta) \bigl( 1 + J M^{-1} ( \delta / \bar{\delta} )^t (2 \bar{\delta})^n \bigr)^{-1} \label{eqn:coding_bound_hy}
\end{IEEEeqnarray}
respectively, where $\psi_N(t; n, \delta) = (\Pr[\operatorname{Bin}(n, \delta) \geq t])^N - (\Pr[\operatorname{Bin}(n, \delta) \geq t + 1])^N$ is the pmf of the minimum of $N$ binomial random variables.
In \cref{fig:bound_comparison}, we illustrate the achievable rates corresponding to the bounds in \eqref{eqn:coding_bound_lpml}--\eqref{eqn:coding_bound_hy} for short blocklengths $n \in \{ 10, 20 \}$, $K$ between 1 and 1024, $\delta = 0.05$ and error tolerance $\varepsilon = 10^{-2}$.

While the baseline performs well for small $K$, its performance saturates.
Intuitively, as $K$ becomes large with $n$ fixed, $\psi_K(t; n, \delta)$ approaches a point mass at $t = 0$; while this effectively eliminates the channel noise, the error probability in \eqref{eqn:coding_bound_bl} is not forced to zero because the shared codebook itself remains a source of errors.
Instead, \eqref{eqn:coding_bound_bl} approaches $(1 + M^{-1} (2 \bar{\delta})^n)^{-1} > 0$, reflecting the fact that when the codebook contains a competing codeword likely to score higher than the true codeword during index selection, it may be erroneously chosen by all decoders simultaneously.
Conversely, taking the limit of \eqref{eqn:coding_bound_lpml} confirms the disjoint scheme has $P_e \to 0$ as $K \to \infty$.
The hybrid strategy chooses $J$ through a 1D search and consistently performs as well or better in both regimes.
Counterintuitively, increasing $n$ from 10 to 20 can decrease the achievable rate for the disjoint and hybrid schemes when $K$ is large.
While the rate asymptotically increases with the blocklength, the benefit of increasing $n$ may be outweighed by the $1/n$ penalty incurred against the codebook diversity gain, as seen in \eqref{eqn:jscc_dispersion_condition} and \eqref{eqn:hybrid_dispersion_condition}, in the regime where $K$ is relatively large while $n$ is small.

\section{Conclusion}\label{sec:conclusion}
We derived one-shot achievability results for a broadcast JSCC problem in a distributed setting where success is determined by the most favorable outcome among $K$ decoders.
We identified two competing ways in which having multiple decoders can reduce the error probability, namely channel diversity, achieved by sharing a codebook and taking multiple independent looks at the channel, and codebook diversity, which requires forcing the decoders' codebooks to be disjoint.
During our analysis, we developed the list PML as an extension of the PML in order to obtain bounds on the error probability.
Moreover, we derived a hybrid coding scheme that optimally balances channel and codebook diversity; numerical results demonstrated that it improves on approaches that rely on one form of diversity alone.

\bibliography{references}

\ifpreprint
\appendices
\crefalias{section}{appendix}
\section{Proof of \Cref{lemma:clml}}\label{sec:lml_proof}
Let us first examine the probability of the complementary event, i.e.\ the matching probability
\begin{IEEEeqnarray}{rCl}
    \Pr[\mathrm{accept}] & = & \Pr[U_p \in \{ U_{qk} \}_{k=1}^{K} \mid U_p] \\
    & = & \sum_{k=1}^{K} \Pr[U_p \in \{ U_{qk} \}_{k=1}^{K}, Z_p = k \mid U_p] \\
    & = & K \Pr[U_p \in \{ U_{qk} \}_{k=1}^{K}, Z_p = 1 \mid U_p]
\end{IEEEeqnarray}
where the last step follows from the symmetry of the construction.
Then, we have that
\begin{IEEEeqnarray}{rCl}
    \IEEEeqnarraymulticol{3}{l}{
        \Pr[U_p \in \{ U_{qk} \}_{k=1}^{K}, Z_p = 1 \mid U_p]
    }\nonumber\\* \quad
    & \geq & \Pr[U_p = U_{q1}, Z_p = 1 \mid U_p] \\
    & = & \Pr[U_p = U_{q1} \mid U_p, Z_p = 1] / K
\end{IEEEeqnarray}
since $Z_p$ is uniform on $\{ 1, \ldots, K \}$ independent of $U_p$.
Using the identity $P_e = 1 - \Pr[\text{accept}]$, we can upper bound the error probability as
\begin{IEEEeqnarray}{c}
    \Pr[U_p \notin \{ U_{qk} \}_{k=1}^{K} \mid U_p] \leq \Pr[U_p \neq U_{q1} \mid U_p, Z_p = 1]. \IEEEeqnarraynumspace \label{eqn:bound}
\end{IEEEeqnarray}
Define $f(u) = p(u) / \mu(u)$ and $g(u) = q(u) / \mu(u)$.
Now let $\tilde{\tau}_i = \tau_i / f(U_i)$.
By the mapping theorem \cite{last2017}, $\{ U_i, \tilde{\tau}_i \}_{i \in \mathbb{N}}$ is a marked Poisson point process with intensity measure $P \times \lambda_{\mathbb{R}_{\geq 0}}$.
Further, let $\tilde{\tau}_p = \min_{i \geq 1} \tilde{\tau}_i$.
Since $K_p = \argmin_{i \geq 1} \tilde{\tau}_i$, we immediately see that $\tilde{\tau}_p = \tilde{\tau}_{K_p}$.
Moreover, since $\tilde{\tau}_p$ is the first arrival time of $\{ U_i, \tilde{\tau}_i \}_{i \in \mathbb{N}}$, we have in addition that $\tilde{\tau}_p \sim \operatorname{Exp}(1)$.
As the mark corresponding to this arrival is $U_p$, we note that $U_p \sim P$ independent of $\tilde{\tau}_p$.

Throughout the rest of the proof, condition on $\tilde{\tau}_p = t$, $U_p = u$ and $Z_p = 1$ unless stated otherwise.
For each $i \neq K_p$, $\tilde{\tau}_i \geq \tilde{\tau}_p = t$.
As shown in \cite{li2021}, the process $\Pi' = \{ U_i, \tilde{\tau}_i \}_{i \neq K_p}$, which excludes the first arrival at time $t$, is then a marked Poisson point process with intensity measure $P \times \lambda_{[t, \infty)}$ independent of $U_p$.
It is also independent of each  $Z_i$.
By the thinning property \cite{last2017}, $\Pi_1' = \{ U_i, \tilde{\tau}_i \}_{i \neq K_p, Z_i = 1}$ is then a marked Poisson point process with intensity measure $(P \times \lambda_{[t, \infty)}) / K$.

To proceed, assume $p(u) > 0$, which holds almost surely since $U_p \sim P$.
If $q(u) = 0$, then the error probability on the left-hand side of \eqref{eqn:clml} is one, since there exists almost surely for each $k$ a point $(U_i, \tau_i)$ for which the score $\tau_i \mu(U_i) / q(U_i)$ is finite and $Z_i = k$, whereas the score for $(U_{K_p}, \tau_{K_p})$ is infinite.
In this case the desired inequality holds trivially.
On the other hand, if $q(u) > 0$, having $U_p \neq U_{q1}$ implies the selected indices are different, i.e.\ $\{ U_p \neq U_{q1} \} \subseteq \{ K_p \neq K_{q1} \}$.
Since the probability of a tie is zero, the probability of an index mismatch is the same as that of seeing
\begin{IEEEeqnarray}{l}
    \min_{i \neq K_p, Z_i = 1} \frac{\tau_i}{g(U_i)} < \frac{\tau_{K_p}}{g(U_{K_p})} \nonumber \\
    \IEEEeqnarraymulticol{1}{r}{
        \qquad \< \iff \min_{i \neq K_p, Z_i = 1} \frac{\tilde{\tau}_i p(U_i)}{q(U_i)} < \frac{\tilde{\tau}_p p(U_p)}{q(U_p)} = \frac{t p(u)}{q(u)}. \IEEEeqnarraynumspace
    }
\end{IEEEeqnarray}
Thus, an upper bound for $\Pr[U_p \neq U_{q1} \mid \tilde{\tau}_p = t, U_p = u, Z_p = 1]$ is the probability of $\Pi_1'$ seeing at least one arrival in the region
\begin{IEEEeqnarray}{c}
    \mathcal{B} = \left\{ (v, s) : s < \frac{t p(u) q(v)}{q(u) p(v)} \right\}.
\end{IEEEeqnarray}
Let the number of arrivals in the region be $N$, which is a Poisson random variable with rate
\begin{IEEEeqnarray}{rCl}
    \alpha & = & ((P \times \lambda_{[t, \infty)}) / K)(\mathcal{B}) \label{eqn:poisson_rate_start} \\
    & = & \frac{1}{K}\int_{\mathcal{U}} \max\biggl\{ 0,  \frac{t p(u) q(v)}{q(u) p(v)} - t \biggr\} p(v) \mu(dv) \\
    & = & \frac{t}{K} \int_{\mathcal{U}} \max\biggl\{ 0, \frac{p(u)}{q(u)} - \frac{p(v)}{q(v)} \biggr\} q(v) \mu(dv) \\
    & \leq & \frac{t}{K} \int_{\mathcal{U}} \frac{p(u) q(v)}{q(u)} \mu(dv) = \frac{t p(u)}{K q(u)}. \label{eqn:poisson_rate_end}
\end{IEEEeqnarray}
Therefore, the probability of the event $\{ N = 0 \}$ is
\begin{IEEEeqnarray}{c}
    \Pr[N = 0 \mid \tilde{\tau}_p = t, U_p = u, Z_p = 1] \geq e^{-t p(u) / K q(u)}. \IEEEeqnarraynumspace
\end{IEEEeqnarray}
As seen earlier, $\tilde{\tau}_p \sim \operatorname{Exp}(1)$ and is independent of $U_p$.
Moreover, $\tilde{\tau}_p$ is also independent of $Z_p$.
As a result,
\begin{IEEEeqnarray}{rCl}
    \Pr[N = 0 \mid U_p = u, Z_p = 1] & \geq & \int_{0}^{\infty} e^{-t} e^{-t p(u) / K q(u)} dt \IEEEeqnarraynumspace \\
    & = & (1 + p(u) / K q(u))^{-1}.
\end{IEEEeqnarray}
Finally, we get
\begin{IEEEeqnarray}{rCl}
    \IEEEeqnarraymulticol{3}{l}{
        \Pr[U_p \neq U_{q1} \mid U_p = u, Z_p = 1]
    }\nonumber\\* \:
    & \leq & \Pr[N > 0 \mid U_p = u, Z_p = 1] \leq 1 - \biggl( 1 + \frac{p(u)}{K q(u)} \biggr)^{-1} \IEEEeqnarraynumspace
\end{IEEEeqnarray}
and recover \eqref{eqn:clml} after substituting into \eqref{eqn:bound}.

\section{Proof of \Cref{lemma:cclml}}\label{sec:conditional_lml_proof}
Instead of proving \cref{lemma:cclml} directly, we instead establish the following generalized version that subsumes the analysis of the baseline and hybrid schemes introduced in \cref{sec:main_results} by allowing the codebook to be partially or fully shared between the decoders.

\begin{lemma}\label{lem:cclml_generalized}
Let $\Pi = \{ U_i, \tau_i \}_{i \in \mathbb{N}}$ be a marked Poisson point process on $\mathcal{U} \times \mathbb{R}_{\geq 0}$ with intensity measure $\mu \times \lambda_{\mathbb{R}_{\geq 0}}$.
Let $L$, $J \in \mathbb{N}$ be such that $K = L J$ and associate with each point of the process $\Pi$ an independent discrete random variable $Z_i$ which is uniformly distributed on $\{ 1, \ldots, J \}$.
Fix a joint distribution $P_{X,U,Y}$ together with a probability kernel $Q_{U \mid Y}$ satisfying $P_{U \mid X}(\cdot \mid X)$, $Q_{U \mid Y}(\cdot \mid Y) \ll \mu$ almost surely, and take $X \sim P_X$ independent of $\Pi$ and each $Z_i$.
For the selection rule, first choose
\begin{IEEEeqnarray}{c}
    K_p = \argmin_{i \geq 1} \frac{\tau_i \mu(U_i)}{p_{U \mid X}(U_i \mid X)}
\end{IEEEeqnarray}
and set $U_p = U_{K_p}$, $Z_p = Z_{K_p}$.
Given $U_p$, let $Y_{1:K} \sim P_{Y \mid X, U}(\cdot \mid X, U_p)$ iid.
We have $(X, U_p, Y_k) \sim P_{X,U,Y}$ for each $k$ and note that the Markov chain condition $Y_{1:K} \leftrightarrow (X, U_p) \leftrightarrow \{ U_i, \tau_i \}_{i \in \mathbb{N}}$ holds by construction.
Next, select
\begin{IEEEeqnarray}{c}
    K_{qk} = \argmin_{i \geq 1, Z_i = \ceil{k/L}} \frac{\tau_i \mu(U_i)}{q_{U \mid Y}(U_i \mid Y_k)}.
\end{IEEEeqnarray}
With $U_{qk} = U_{K_{qk}}$, we have
\begin{IEEEeqnarray}{rCl}
    \IEEEeqnarraymulticol{3}{l}{
        \Pr[U_p \notin \{ U_{qk} \}_{k=1}^{K} \mid X, U_p, \: Y_{1:K}]
    }\nonumber\\* \quad
    & \leq & 1 - \sum_{j=1}^{J}\biggl( J + \min_{(j-1)L + 1 \leq k \leq jL} \frac{p_{U \mid X}(U_p \mid X)}{q_{U \mid Y}(U_p \mid Y_k)} \biggr)^{-1}. \IEEEeqnarraynumspace \label{eqn:cclml_generalized}
\end{IEEEeqnarray}
Note that \cref{lemma:cclml} can be recovered from \eqref{eqn:cclml_generalized} by setting $L = 1$.
\end{lemma}
\begin{IEEEproof}
For convenience, let $\pi(j) = \{ (j-1)L + 1, \ldots, jL \}$ for $1 \leq j \leq J$.
We start by expanding the conditional acceptance probability,
\begin{IEEEeqnarray}{rCl}
    \IEEEeqnarraymulticol{3}{l}{
        \Pr[\mathrm{accept}]
    }\nonumber\\* \quad
    & = & \Pr[U_p \in \{ U_{qk} \}_{k=1}^{K} \mid X, U_p, Y_{1:K}] \\
    & = & \sum_{j=1}^{J} \Pr[U_p \in \{ U_{qk} \}_{k=1}^{K}, Z_p = j \mid X, U_p, Y_{1:K}] \\
    & \geq & \sum_{j=1}^{J} \Pr[U_p \in \{ U_{qk} \}_{k \in \pi(j)}, Z_p = j \mid X, U_p, Y_{1:K}] \\
    & = & \sum_{j=1}^{J} \Pr[U_p \in  \{ U_{qk} \}_{k \in \pi(j)} \mid X, U_p, Y_{1:K}, Z_p = j] / J \IEEEeqnarraynumspace \label{eqn:accept_prob}
\end{IEEEeqnarray}
since $Z_p$ is uniform on $\{ 1, \ldots, J \}$ independent of $X$, $U_p$ and $Y_{1:K}$.
The analysis will be the same for each $j$, so let us focus on $j = 1$ and consider
\begin{IEEEeqnarray}{c}
    \Pr[U_p \in \{ U_{qk} \}_{k=1}^{L} \mid X, U_p, Y_{1:K}, Z_p = 1]
\end{IEEEeqnarray}
where we have used the fact that $\pi(1) = \{ 1, \ldots, L \}$.
Let $f(u \mid X) = p_{U \mid X}(u \mid X) / \mu(u)$ and $g(u \mid Y_k) = q_{U \mid Y}(u \mid Y_k) / \mu(u)$.
Let $\tilde{\tau}_i = \tau_i / f(U_i \mid X)$.
By the mapping theorem \cite{last2017}, $\{ U_i, \tilde{\tau}_i \}_{i \in \mathbb{N}}$ is then a marked Poisson point process with intensity measure $P_{U \mid X}(\cdot \mid X) \times \lambda_{\mathbb{R}_{\geq 0}}$.
Let $\tilde{\tau}_p = \min_{i \geq 1} \tilde{\tau}_i$.
We immediately see that $\tilde{\tau}_p = \tilde{\tau}_{K_p}$.
Given $X$, we also have that $\tilde{\tau}_p \sim \operatorname{Exp}(1)$ and $U_p \sim P_{U \mid X}(\cdot \mid X)$ independent of $\tilde{\tau}_p$, similar to what was seen during the proof of \cref{lemma:clml}.
This confirms that $(X, U_p, Y_k) \sim P_{X,U,Y}$ for each $k$.

In what follows, condition on $\tilde{\tau}_p = t$, $U_p = u$, $Z_p = 1$ and $Y_{1:K}$.
Mirroring the arguments used to prove \cref{lemma:clml}, $\tilde{\tau}_i \geq \tilde{\tau}_p = t$ for each $i \neq K_p$ and therefore, by excluding the first arrival at time $t$, $\Pi' = \{ U_i, \tilde{\tau}_i \}_{i \neq K_p}$ becomes a marked Poisson point process with intensity measure $P_{U \mid X}(\cdot \mid X) \times \lambda_{[t, \infty)}$ independent of $U_p$ and each $Z_i$.
By the thinning property, $\Pi_1' = \{ U_i, \tilde{\tau}_i \}_{i \neq K_p, Z_i = 1}$ is then a marked Poisson point process with intensity measure $(P_{U \mid X}(\cdot \mid X) \times \lambda_{[t, \infty)}) / J$.
By the Markov chain condition noted earlier, both $\Pi'$ and $\Pi_1'$ are also conditionally independent of $Y_{1:K}$ given $X$ and $U_p$.
As in the unconditional case, assume $p_{U \mid X}(u \mid X) > 0$.
First consider that $q_{U \mid Y}(u \mid Y_k) > 0$ for at least one $k \in \pi(1)$.
Selecting the same indices implies the selected values are the same, and consequently
\begin{IEEEeqnarray}{c}
    \{ K_p = K_{q{k^\ast}} \} \! \subseteq \! \{ K_p \in \{ K_{qk} \}_{k=1}^{L} \} \! \subseteq \! \{ U_p \in \{ U_{qk} \}_{k=1}^{L} \} \IEEEeqnarraynumspace
\end{IEEEeqnarray}
for any choice of $1 \leq k^\ast \leq L$.
Similar to what was seen in the proof of \cref{eqn:clml}, the probability of the event $\{ K_p \neq K_{qk^\ast} \}$ is the same as that of seeing
\begin{IEEEeqnarray}{c}
    \min_{i \neq K_p, Z_i = 1} \frac{\tilde{\tau}_i p_{U \mid X}(U_i \mid X)}{q_{U \mid Y}(U_i \mid Y_{k^\ast})} < \frac{\tilde{\tau}_p p_{U \mid X}(U_p \mid X)}{q_{U \mid Y}(U_p \mid Y_{k^\ast})}.
\end{IEEEeqnarray}
Then, a lower bound for $\Pr[U_p \in \{ U_{qk} \}_{k=1}^{L} \mid \tilde{\tau}_p = t, X, U_p = u, Y_{1:K}, Z_p = 1]$ is given by the probability of $\Pi_1'$ seeing no arrivals in the region
\begin{IEEEeqnarray}{c}
    \mathcal{B}_{k^\ast} = \left\{ (v, s) : s < \frac{t p_{U \mid X}(u \mid X) q_{U \mid Y}(v \mid Y_{k^\ast})}{q_{U \mid Y}(u \mid Y_{k^\ast}) p_{U \mid X}(v \mid X)} \right\}. \IEEEeqnarraynumspace
\end{IEEEeqnarray}
Let the number of arrivals in the region be $N_{k^\ast}$.
Repeating the sequence of steps in \eqref{eqn:poisson_rate_start}--\eqref{eqn:poisson_rate_end}, $N_{k^\ast}$ is a Poisson random variable with rate $\alpha \leq t p_{U \mid X}(u \mid X) / J q_{U \mid Y}(u \mid Y_{k^\ast})$.
The probability of zero arrivals is then
\begin{IEEEeqnarray}{rCl}
    \IEEEeqnarraymulticol{3}{l}{
        \Pr[N_{k^\ast} = 0 \mid \tilde{\tau}_p = t, X, U_p = u, Y_{1:K}, Z_p=1]
    }\nonumber\\* \quad
    & \geq & e^{-t p_{U \mid X}(u \mid X) / J q_{U \mid Y}(u \mid Y_{k^\ast})}.
\end{IEEEeqnarray}
As seen earlier, $\tilde{\tau}_p \sim \operatorname{Exp}(1)$ independent of $U_p$ after conditioning on $X$.
Moreover, $\tilde{\tau}_p$ is conditionally independent of $Y_{1:K}$ given $X$ and $U_p$ and is also independent of $Z_p$.
Therefore, we can integrate out the exponential density of $\tilde{\tau}_p$ to get
\begin{IEEEeqnarray}{rCl}
    \IEEEeqnarraymulticol{3}{l}{
        \Pr[N_{k^\ast} = 0 \mid X, U_p = u, Y_{1:K}, Z_p = 1]
    }\nonumber\\* \quad
    & \geq & (1 + p_{U \mid X}(u \mid X) / J q_{U \mid Y}(u \mid Y_{k^\ast}))^{-1}.
\end{IEEEeqnarray}
We now have
\begin{IEEEeqnarray}{rCl}
    \IEEEeqnarraymulticol{3}{l}{
        \Pr[U_p \in \{ U_{qk} \}_{k=1}^{L} \mid X, U_p = u, Y_{1:K}, Z_p = 1]
    }\nonumber\\* \quad
    & \geq & \Pr[N_{k^\ast} = 0 \mid X, U_p = u, Y_{1:K}, Z_p = 1] \\
    & \geq & (1 + p_{U \mid X}(u \mid X) / J q_{U \mid Y}(u \mid Y_{k^\ast}))^{-1}.
\end{IEEEeqnarray}
Since the inequality holds simultaneously for any choice of $k^\ast$, we can choose $k^\ast = \argmin_{1 \leq k \leq L} p_{U \mid X}(u \mid X) / q_{U \mid Y}(u \mid Y_k)$ to see that
\begin{IEEEeqnarray}{rCl}
    \IEEEeqnarraymulticol{3}{l}{
        \Pr[U_p \in \{ U_{qk} \}_{k=1}^{L} \mid X, U_p = u, Y^K, Z_p = 1]
    }\nonumber\\* \quad
    & \geq & \biggl( 1 + \frac{1}{J} \min_{1 \leq k \leq L} \frac{p_{U \mid X}(u \mid X)}{q_{U \mid Y}(u \mid Y_k)} \biggr)^{-1}.\label{eqn:cclml_ineq}
\end{IEEEeqnarray}
On the other hand, if $q_{U \mid Y}(u \mid Y_k) = 0$ for all $1 \leq k \leq L$, the conditional probability of the event $\{ U_p \in \{ U_{qk} \}_{k=1}^{L} \}$ is zero since for each $k$ there exists almost surely a point $(U_i, \tau_i)$ for which the score $\tau_i \mu(U_i) / q_{U \mid Y}(U_i \mid Y_k)$ is finite and $Z_i = 1$, whereas the score for $(U_{K_p}, \tau_{K_p})$ is infinite.
The bound in \eqref{eqn:cclml_ineq} also holds in this case.
After generalizing \eqref{eqn:cclml_ineq} to $Z_p = j$ for any $j$ and substituting into \eqref{eqn:accept_prob}, we obtain
\begin{IEEEeqnarray}{rCl}
    \IEEEeqnarraymulticol{3}{l}{
        \Pr[\mathrm{accept}]
    }\nonumber\\* \quad
    & \geq & \sum_{j=1}^{J} \biggl( J + \min_{(j-1)L + 1 \leq k \leq jL} \frac{p_{U \mid X}(U_p \mid X)}{q_{U \mid Y}(U_p \mid Y_k)} \biggr)^{-1}. \IEEEeqnarraynumspace
\end{IEEEeqnarray}
Taking the complement yields \eqref{eqn:cclml_generalized}.
\end{IEEEproof}

\section{Error Probability of the Hybrid Scheme}\label{sec:hybrid_proof}
In this section, we provide a proof of \eqref{eqn:hybrid_bound} via \cref{lem:cclml_generalized}.
By setting $J = 1$, this also serves to show the bound in \eqref{eqn:baseline_bound} for the baseline scheme's performance.
Following the approach used in the proof of \cref{thm:oneshot_jscc}, let the codebook be a marked Poisson point process $\Pi = \{ (X_i, Z_i), \tau_i \}_{i \in \mathbb{N}}$ on $\mathcal{X} \times \mathcal{Z} \times \mathbb{R}_{\geq 0}$ with intensity measure $P_X \times P_Z \times \lambda_{\mathbb{R}_{\geq 0}}$ independent of the source $W$.
Associate with each point a label $C_i$ uniform on $\{ 1, \ldots, J \}$ independent of $\Pi$.
The encoder uses the modified conditional distribution $P_{\tilde{Z} \mid W}$ as in \eqref{eqn:modified_encoder_dist} and selects its index $K_p$ using \eqref{eqn:thm1_encoder_selection}.
Meanwhile, decoder $k$ selects $K_{qk}$ using \eqref{eqn:hybrid_decoder_selection}.
As before, let $X = X_{K_p}$, $\tilde{Z} = Z_{K_p}$, $\hat{Z}_k = Z_{K_{qk}}$ and $\hat{X}_{k} = X_{K_{qk}}$.
The channel outputs observed by the decoders follow $Y_{1:K} \sim P_{Y \mid X}(\cdot \mid X)$ iid.
We have $(W, \tilde{Z}, X, Y_k) \sim P_W P_{\tilde{Z} \mid W} \times P_X P_{Y \mid X}$ for each $k$ by \cref{lem:cclml_generalized}.
Additionally, define $\pi(j) = \{ (j-1)L + 1, \ldots, jL \}$ for $1 \leq j \leq J$.
Then,
\begin{IEEEeqnarray}{rCl}
    \IEEEeqnarraymulticol{3}{l}{
        \Pr[\cap_{k=1}^{K} \{ d(W, \hat{Z}_k) > D \}]
    }\nonumber\\* \:
    & \leq & \Pr[\rho(W) = 0] + \E[\1\{ \rho(W) > 0 \} \nonumber \\
    &&\< \times \Pr[(X, \tilde{Z}) \notin \{ (\hat{X}_k, \hat{Z}_k) \}_{k=1}^{K} \mid W, X, \tilde{Z}, Y_{1:K}]] \\
    & \stackrel{\text{(a)}}{\leq} & \Pr[\rho(W) = 0] + \E\biggl[\1\{ \rho(W) > 0 \} \nonumber \\
    &&\< \times \biggl( 1 - \sum_{j=1}^{J} \biggl( J + \min_{k \in \pi(j)} \frac{p_X(X) p_{\tilde{Z} \mid W}(\tilde{Z} \mid W)}{p_{X \mid Y}(X \mid Y_k) p_Z(\tilde{Z})} \biggr)^{-1} \biggr)\biggr] \\
    & = & \Pr[\rho(W) = 0] + \E\biggl[\1\{ \rho(W) > 0 \} \nonumber \\
    &&\< \times \biggl(1 - \sum_{j=1}^{J} \Bigl(J + \rho(W)^{-1} \min_{k \in \pi(j)} 2^{-\iota_{X;Y}(X;Y_k)}\Bigr)^{-1}\biggr)\biggr] \IEEEeqnarraynumspace \\
    & \stackrel{\text{(b)}}{=} & 1 - J\E\Bigl[ \Bigl( J + \rho(W)^{-1} \min_{1 \leq k \leq L} 2^{-\iota_{X; Y}(X; Y_k)} \Bigr)^{-1} \Bigr] \\
    & = & \E\Bigl[ \Bigl( 1 + J P_Z(\mathcal{B}_D(W)) \max_{1 \leq k \leq L} 2^{\iota_{X;Y}(X;Y_k)} \Bigr)^{-1} \Bigr]
\end{IEEEeqnarray}
where $(W, X, Y_k) \sim P_W \times P_X P_{Y \mid X}$ for each $k$, (a) is by \cref{lem:cclml_generalized} and (b) uses the fact that the $Y_k$'s are identically distributed given $X$ and the identity $\E[\1\{ \rho(W) > 0 \}] = \Pr[\rho(W) > 0]$. \hfill \IEEEQED{}

\section{Proof of \Cref{thm:oneshot_jscc_dispersion}}\label{sec:dispersion_proof}
The approach used to prove \cref{thm:oneshot_jscc_dispersion} is similar that used in \cite[App.~D]{li2021}.
We make use of the following lemma from \cite{kostina2012} which deals with the probability mass inside the distortion ball.
As mentioned in \cref{sec:second_order_analysis}, we assume throughout that all the conditions required by the lemma hold.

\begin{lemma}[{\hspace{1sp}\cite[Lem.~2]{kostina2012}}]\label{lem:kostina2012}
Let $W^m \sim P_W$ iid.
There exist constants $\alpha$, $\beta$, $\gamma$ such that for $m$ sufficiently large,
\begin{IEEEeqnarray}{l}
    \Pr\biggl[ -{\log} P_{Z^m}(\mc{B}_D(W^m)) \nonumber \\
    \IEEEeqnarraymulticol{1}{r}{
        \qquad \leq \sum_{i=1}^{m} \jmath_W(W_i, D) + \alpha \log m + \beta \biggr] \geq 1 - \frac{\gamma}{\sqrt{m}}
    }\label{eqn:kostina2012}
\end{IEEEeqnarray}
where $P_{Z^m} = P_Z^m$, provided the following conditions hold:
\begin{enumerate}
    \item $\inf\{ x \geq 0 : R(x) < \infty \} < D < \inf_{z \in \mc{Z}} \E[d(W,z)]$.
    \item The infimum in $R(D)$ is attained by a unique $P_{Z \mid W}$.
    \item There exists a finite set $\tilde{\mathcal{Z}} \subseteq \mathcal{Z}$ which satisfies $\E[\min_{z \in \tilde{\mathcal{Z}}} d(W,z)] < \infty$.
    \item $\E_{(Z,W) \sim P_Z \times P_W}[d(W,Z)^9] < \infty$.
\end{enumerate}
For convenience, define
\begin{IEEEeqnarray}{c}
    \Omega = \sum_{i=1}^{m} \jmath_W(W_i, D) + \alpha \log m + \beta.
\end{IEEEeqnarray}
Examining the ensemble error event for the transmission, its probability $P_e$ corresponds to
\begin{IEEEeqnarray}{rCl}
    \IEEEeqnarraymulticol{3}{l}{
        \Pr[\cap_{k=1}^{K} \{ d(W^m, \hat{Z}^m_k) > D \}]
    }\nonumber\\* \:
    & \stackrel{\text{(a)}}{\leq} & \E\bigl[ \bigl( 1 + K P_{Z^m}(\mathcal{B}_D(W^m)) 2^{\iota_{X^n;Y^n}(X^n;Y^n)} \bigr)^{-1} \bigr] \\
    & \leq & \Pr[-{\log} P_{Z^m}(\mathcal{B}_D(W^m)) > \Omega] \nonumber \\
    &&\< + \E\bigl[ \bigr(1 + K 2^{\iota_{X^n;Y^n}(X^n; Y^n) - \Omega} \bigl)^{-1} \bigr] \\
    & \stackrel{\text{(b)}}{\leq} & \frac{\gamma}{\sqrt{m}} + \E\bigl[ \bigr( 1 + K 2^{\iota_{X^n;Y^n}(X^n; Y^n) - \Omega} \bigr)^{-1} \bigr] \\
    & = & \frac{\gamma}{\sqrt{m}} + \E\bigl[ \bigl(1 + 2^{\iota_{X^n;Y^n}(X^n; Y^n) - \Omega + \log K} \bigr)^{-1} \bigr]. \IEEEeqnarraynumspace
\end{IEEEeqnarray}
where (a) is by \cref{thm:oneshot_jscc}, in particular applying \eqref{eqn:oneshot_jscc} to $W^m$ and the concatenated channel $P_{Y^n \mid X^n}$, and (b) is by \cref{lem:kostina2012}.
We further split the expectation based on the event
\begin{IEEEeqnarray}{c}
    \{ 2^{\iota_{X^n;Y^n}(X^n; Y^n) - \Omega + \log K} \geq \sqrt{n} \}
\end{IEEEeqnarray}
which gives
\begin{IEEEeqnarray}{rCl}
    P_e & \leq & \frac{\gamma}{\sqrt{m}} + \frac{1}{\sqrt{n}} + \Pr\biggl[ 2^{\Omega - \iota_{X^n;Y^n}(X^n; Y^n) - \log K} > \frac{1}{\sqrt{n}} \biggr] \IEEEeqnarraynumspace \label{eqn:hybrid_proof_start} \\
    & = & \frac{\gamma}{\sqrt{m}} + \frac{1}{\sqrt{n}} + \Pr\biggl[ A < -nC + m R(D) \nonumber \\
    && +\> \alpha \log m + \frac{1}{2} \log n + \beta - \log K \biggr] \label{eqn:prob_term}
\end{IEEEeqnarray}
where we have introduced
\begin{IEEEeqnarray}{c}
    A = \sum_{i=1}^{n} (\iota_{X;Y}(X_i;Y_i) - C) - \sum_{i=1}^{m} (\jmath_W(W_i,D) - R(D)). \IEEEeqnarraynumspace
\end{IEEEeqnarray}
Note that $\E[A] = 0$ and $\Var(A) = nV + m \mathcal{V}(D)$.
The probability term in \eqref{eqn:prob_term} can be bounded as
\begin{IEEEeqnarray}{rCl}
    \IEEEeqnarraymulticol{3}{l}{
         \Pr\biggl[ A < -nC + m R(D) + \alpha \log m + \frac{1}{2} \log n + \beta - \log K \biggr]
    }\nonumber\\* \:
    & \stackrel{\text{(c)}}{\leq} & \Pr\biggl[ A \! \leq \! -\sqrt{nV + m \mathcal{V}(D)} Q^{-1}\biggl( \varepsilon - \frac{\eta}{\sqrt{\min\{ n, m \}}} \biggr) \biggr] \IEEEeqnarraynumspace \\
    & = & \Pr\biggl[ \frac{A}{\sqrt{nV + m\mathcal{V}(D)}} \leq -Q^{-1}\biggl( \varepsilon - \frac{\eta}{\sqrt{\min\{ n, m \}}} \biggr) \biggr] \\
    & \stackrel{\text{(d)}}{\leq} & \Phi\biggl( -Q^{-1}\biggl( \varepsilon - \frac{\eta}{\sqrt{\min\{ n, m \}}} \biggr) \biggr) + \frac{B}{\sqrt{\min\{ n, m \}}} \\
    & = & \varepsilon + \frac{B - \eta}{\sqrt{\min\{ n, m \}}}
\end{IEEEeqnarray}
where (c) is due to the condition in \eqref{eqn:jscc_dispersion_condition} and (d) is by the Berry-Esseen theorem \cite{feller1971} with $B$ being a positive constant.
Then,
\begin{IEEEeqnarray}{c}
    P_e \leq \frac{\gamma}{\sqrt{m}} + \frac{1}{\sqrt{n}} + \varepsilon + \frac{B - \eta}{\sqrt{\min\{ n, m \}}}. \IEEEeqnarraynumspace
\end{IEEEeqnarray}
Letting $\eta = B + \gamma + 1$ then shows $P_e \leq \varepsilon$. \hfill \IEEEQED{}

\end{lemma}

\section{Proof of \Cref{thm:hybrid_second_order}}\label{sec:hybrid_dispersion_proof}
The start of the proof mirrors that of the basic second-order result in \cref{thm:oneshot_jscc_dispersion}.
In particular, applying the steps in \cref{sec:dispersion_proof} up to \eqref{eqn:hybrid_proof_start} on the hybrid JSCC bound in \eqref{eqn:hybrid_bound} shows that
\begin{IEEEeqnarray}{rCl}
    P_e & \leq & \frac{\gamma}{\sqrt{m}} + \frac{1}{\sqrt{n}} \nonumber \\
    && \< + \Pr\biggl[ 2^{\Omega - \max_{i \leq k \leq L} \iota_{X^n;Y^n}(X^n; Y_k^n) - \log J} > \frac{1}{\sqrt{n}} \biggr] \IEEEeqnarraynumspace
\end{IEEEeqnarray}
where $\Omega$ is defined as before.
Let us therefore focus on bounding the probability
\begin{IEEEeqnarray}{rCl}
    \IEEEeqnarraymulticol{3}{l}{
        \Pr\biggl[ \max_{1 \leq k \leq L} \iota_{X^n;Y^n}(X^n; Y_k^n) - \sum_{i=1}^{m} \jmath_{W}(W_i, D)
    }\nonumber\\* \:
    && \< < \alpha \log m + \frac{1}{2} \log n + \beta - \log J \biggr] \\
    & = & \Pr\biggl[ A_{\mathrm{max}} \sqrt{n + m} \nonumber \\
    && \< < -nC + mR(D) + \alpha \log m + \frac{1}{2} \log n + \beta - \log J \biggr] \label{eqn:hybrid_dispersion_ineq} \IEEEeqnarraynumspace
\end{IEEEeqnarray}
where $A_{\mathrm{max}}$ is defined to be the maximum coordinate of the length-$L$ random vector
\begin{IEEEeqnarray}{c}
    A = \frac{1}{\sqrt{n + m}} \sum_{i=1}^{n+m} A_i
\end{IEEEeqnarray}
with each $A_i$ being given by
\begin{IEEEeqnarray}{c}
    A_i = \begin{cases}
        \begin{bsmallmatrix} \iota_{X;Y}(X_i; Y_{1i}) - C & \cdots & \iota_{X;Y}(X_i; Y_{Li}) - C \end{bsmallmatrix}^T, \\
        \qquad 1 \leq i \leq n \\
        \begin{bsmallmatrix} R(D) - \jmath_W(W_{i-n}, D) & \cdots & R(D) - \jmath_W(W_{i-n}, D) \end{bsmallmatrix}^T, \\
        \qquad n + 1 \leq i \leq n + m.
    \end{cases} \IEEEeqnarraynumspace
\end{IEEEeqnarray}
Note that $\E[A_i] = 0$ for all $i$.
We now define the sample covariance matrix
\begin{IEEEeqnarray}{c}
    \Sigma = \frac{1}{n + m} \sum_{i=1}^{n+m} \E[A_i A_i^T].
\end{IEEEeqnarray}
For $n + 1 \leq i \leq n + m$, we have that $\E[A_i A_i^T] = \mathcal{V}(D) 1_{L \times L}$.
For $1 \leq i \leq n$, the diagonal elements of $\E[A_i A_i^T]$ are equal to $\Var(\iota_{X;Y}(X_i; Y_{ki})) = V$.
For the off-diagonal elements, noting that that $Y_{ki}$ and $Y_{ji}$ are conditionally independent given $X_i$ when $k \neq j$, we can use the law of total covariance to find the off-diagonal term as
\begin{IEEEeqnarray}{rCl}
    \IEEEeqnarraymulticol{3}{l}{
        \Cov(\iota_{X;Y}(X_i; Y_{ki}), \iota_{X;Y}(X_i; Y_{ji}))
    }\nonumber\\* \,
    & = & \E[\Cov(\iota_{X;Y}(X_i; Y_{ki}), \iota_{X;Y}(X_i; Y_{ji}) \mid X_i)] \nonumber \\
    && + \Cov(\E[\iota_{X;Y}(X_i; Y_{ki}) \mid X_i], \E[\iota_{X;Y}(X_i; Y_{ji}) \mid X_i]) \IEEEeqnarraynumspace \\
    & = & \Cov(\E[\iota_{X;Y}(X_i; Y_{ki}) \mid X_i], \E[\iota_{X;Y}(X_i; Y_{ji}) \mid X_i]) \\
    & = & \Var(\E[\iota_{X;Y}(X; Y) \mid X])
\end{IEEEeqnarray}
where $(X, Y) \sim P_X P_{Y \mid X}$.
Note further that
\begin{IEEEeqnarray}{rCl}
    \E[\iota_{X;Y}(X; Y) \mid X] & = & \E_{Y \sim P_{Y \mid X}(\cdot \mid X)} \! \biggl[ \log\frac{p_{Y \mid X}(Y \mid X)}{p_Y(Y)} \biggr] \IEEEeqnarraynumspace \\
    & = & D_{\mathrm{KL}}(P_{Y \mid X}(\cdot \mid X) \Vert P_Y).
\end{IEEEeqnarray}
Therefore, the full sample covariance matrix becomes
\begin{IEEEeqnarray}{c}
    \Sigma = \frac{\sigma_1^2 1_{L \times L} + \sigma^2_2 I}{n + m}
\end{IEEEeqnarray}
where $\sigma_1^2 = n \tilde{V} + m \mathcal{V}(D)$ and $\sigma_2^2 = n(V - \tilde{V})$.

Let $F(t; \Sigma)$ be the cdf of the maximum coordinate of a Gaussian random vector with mean zero and covariance matrix $\Sigma$ and let $F^{-1}(t; \Sigma)$ be the associated quantile function.
By applying the Chernozhukov-Chetverikov-Kato bound \cite{chernozhukov2013} after making suitable moment assumptions, see \cite{chernozhukov2013}, we have that
\begin{IEEEeqnarray}{c}
    \Pr[A_{\mathrm{max}} \leq t] \leq F(t; \Sigma) + B (n+m)^{-c}
\end{IEEEeqnarray}
for any $t \in \mathbb{R}$, where $B$, $c > 0$ are constants.
Applying the condition in \eqref{eqn:hybrid_dispersion_condition} gives
\begin{IEEEeqnarray}{rCl}
    P_e & \leq & \frac{\gamma}{\sqrt{m}} + \frac{1}{\sqrt{n}} + B(n+m)^{-c} \nonumber \\
    && \< + F\biggl( F^{-1}\biggl( \varepsilon - \frac{\eta}{( \min\{ n, m \} )^{ \min\{ c, 1/2 \}}}; \Sigma \biggr) \biggr) \IEEEeqnarraynumspace \\
    & \leq & \frac{\gamma + 1 - \eta + B}{( \min\{ n, m \} )^{\min\{ c, 1/2 \}}} + \varepsilon.
\end{IEEEeqnarray}
Letting $\eta = B + \gamma + 1$ completes the proof. \hfill \IEEEQED{}
\fi

\end{document}